\documentclass[prb,superscriptaddress,showpacs,reprint]{revtex4-1}

\usepackage[utf8]{inputenc}
\usepackage{amsmath,amsfonts,amssymb}
\usepackage{pifont}
\usepackage{rotating}
\usepackage{graphicx}    
\usepackage{epstopdf}
\usepackage{color}
\usepackage{soul}
\newcommand{\etal}{\emph{et al.} }
\usepackage{hyperref}
\definecolor{dark-red}{rgb}{0.9,0.15,0.15}
\definecolor{dark-blue}{rgb}{0.15,0.15,0.4}
\definecolor{medium-blue}{rgb}{0,0,0.5}
\hypersetup{
    colorlinks, linkcolor={black},
    citecolor={dark-blue}, urlcolor={medium-blue}
}

\begin{document}
	
	\title{Demonstration of defect-defect ferromagnetic coupling in Gd doped GaN epitaxial films: A polarization selective magneto-photoluminescence study}
	
	\author{Rajendra K. Saroj}
	\affiliation{Department of Physics, Indian Institute of Technology Bombay, Mumbai 400076, India}
	
	\author{Preetha Sarkar}
	\affiliation{Department of Physics, Indian Institute of Technology Bombay, Mumbai 400076, India}
	
	\author{Swarup Deb}
	\affiliation{Department of Physics, Indian Institute of Technology Bombay, Mumbai 400076, India}
	
	\author{S. Dhar}
	\email{dhar@phy.iitb.ac.in}
	\affiliation{Department of Physics, Indian Institute of Technology Bombay, Mumbai 400076, India}

	\begin{abstract}
		Magnetic field dependent polarization selective photoluminescence(PL) study has been carried out  at 1.5~K on Gd-doped GaN epitaxial layers grown on c-SiC substrates by molecular beam epitaxy technique. It has been found that the incorporation of Gd in GaN leads to the generation of three types of donor like defects that result in neutral donor bound excitonic features in low temperature PL. The study reveals that the rate of spin-flip scattering for all the three excitonic features becomes almost $B$-independent suggesting that these signals must be stemming from defects which are ferromagnetically coupled with each other. This is further confirmed by the study carried out on a GaN sample co-doped with Si and Gd, where defects are found to be ferromagnetically coupled, while Si-donors do not show any involvement in coupling.  
	\end{abstract}
	\maketitle

	Gadolinium (Gd)-doped GaN (GaN:Gd) remains to be one of the active areas in physics for exhibiting certain intriguing magnetic properties. Ferromagnetism far above room temperature even with doping concentration as low as $\approx$ 10$^{15}$ cm$^{-3}$ as well as several orders of magnitude larger effective magnetic moment per Gd ion as compared to that of a bare Gd$^{3+}$ ion (8~$\mu_B$)  have been observed in GaN:Gd epitaxial layers\cite{dhar1a, dhar1b}. Element specific magnetic studies on these layers show a very small polarization for Ga and paramagnetism for Gd ions, indicating that the magnetism does not solely arise from Gd itself\cite{ney}. Since then, a large volume of work in this field has revealed ferromagnetic like behavior not only in GaN:Gd\cite{hite1,rizzi,asahi} but also in other rare-earth doped GaN\cite{hite2, sun}. Similar magnetic behavior has also been reported in Gd\cite{dhar2,khaderabad} and Dy\cite{dy-GaN}-implanted GaN layers. Effective magnetic moment per Gd ion, which is reported to be larger in Gd implanted GaN layers,\cite{dhar2} shows a reduction upon annealing\cite{khaderabad}, suggesting a defect origin of the magnetism. Formation of multiple type of defects due to Gd incorporation have been demonstrated in molecular beam epitaxy (MBE) grown GaN:Gd layers\cite{mishra1, mishra2,rizzi}. In these samples, Mishra et al. have also found a connection between the magnetism and the density of certain defects that results in a low temperature photoluminescence(PL) peak at $\approx$ 3.05~eV\cite{mishra2, mishra3}. However, the microscopic origin of the defects and their involvement in establishing long range magnetic ordering is still unclear. Theoretical studies have shown that certain defects, such as Ga-vacancies($V_{Ga}$),\cite{gohda, dev, liu, thiess} N-interstitials($N_{i}$)\cite{mitra},  O-interstitials($O_{i}$)\cite{mitra} as well as nitrogen vacancy($V_{N}$)-Ga vacancy complexes \cite{V_Ga,V_GaN}, which possess magnetic moment, can account for the large effective magnetic moment per Gd ion and explain ferromagnetism in this material.  It is noteworthy that ferromagnetism above room temperature has been observed in semiconductors such as HfO, ZnO, TiO$_2$, In$_2$O$_3$, where atoms with partially filled $d$ or $f$ shells are not present at all\cite{d0fm}. Though, crystalline defects are predicted to be the reason for ferromagnetism in these semiconductors, there is no experimental report, which directly evidences coupling between defects.   
	
	Here, we have carried out a magnetic field dependent polarization selective photoluminescence(PL) study at 1.5~K on several Gd-doped GaN epitaxial layers grown on c-6H SiC substrates by MBE. The study reveals that the incorporation of Gd in GaN leads to the generation of three types of donor like defects, all of which give rise to neutral donor bound excitonic transitions in low temperature PL. It has been shown that the dependence of spin-flip scattering rate on the magnetic field ($B$) for a given excitonic transition carries the information about the magnetic coupling of the associated defects. Spin-flip scattering rates for all the three excitonic features are found to be almost $B$-independent for samples with Gd concentration more than $\approx$ 10$^{17}$~cm$^{-3}$ suggesting that these signals must be stemming from the defects which are ferromagnetically coupled with each other. 
	
	\begin{table}
			\centering
			\caption{Layer thickness $t$, concentration [measured by secondary ion mass spectrometry(SIMS)] of Gd (N$_{Gd}$), Si (N$_{Si}$),  saturation magnetizations M$_S$ recorded at 2~K and 300~K for the samples. D* represents sample D after annealing.}
			\begin{tabular}{c|c|c|c|c|c} 
				\hline
				\hline
				{\small Sample}  &  {\small $t$} & {\small N$_{Gd}$}         & {\small N$_{Si}$}     &  {\small M$_S$ (2~K)}    & {\small M$_S$ (300~K)}\\ 
				& {\small(nm)}& {\small(cm$^{-3}$)}        &  {\small(cm$^{-3}$)} & {\small(emu/cm$^{3}$)} & {\small(emu/cm$^{3}$)} \\ \hline
				{\small C}       & {\small 700} & {\small 6$\times$10$^{16}$} & {\small 0}            & {\small 0.52}             &  {\small 0.41}  \\ 
				{\small D}       & {\small 700} & {\small 2.45$\times$10$^{17}$} & {\small 0}         & {\small 0.8}             & {\small 0.54}  \\ 
			    {\small D*}       & {\small 700} & {\small 2.45$\times$10$^{17}$} & {\small 0}         & {\small -}             & {\small 0.22}  \\ 	
				{\small E}       & {\small 200} & {\small 1$\times$10$^{18}$} & {\small 2$\times$10$^{19}$} &   {\small 4}        &    {\small 2.5} \\ 
				\hline 
				\hline
			\end{tabular}
			\label{tab1}
	\end{table}

\begin{figure}
		\centering
		\includegraphics[width=1\linewidth]{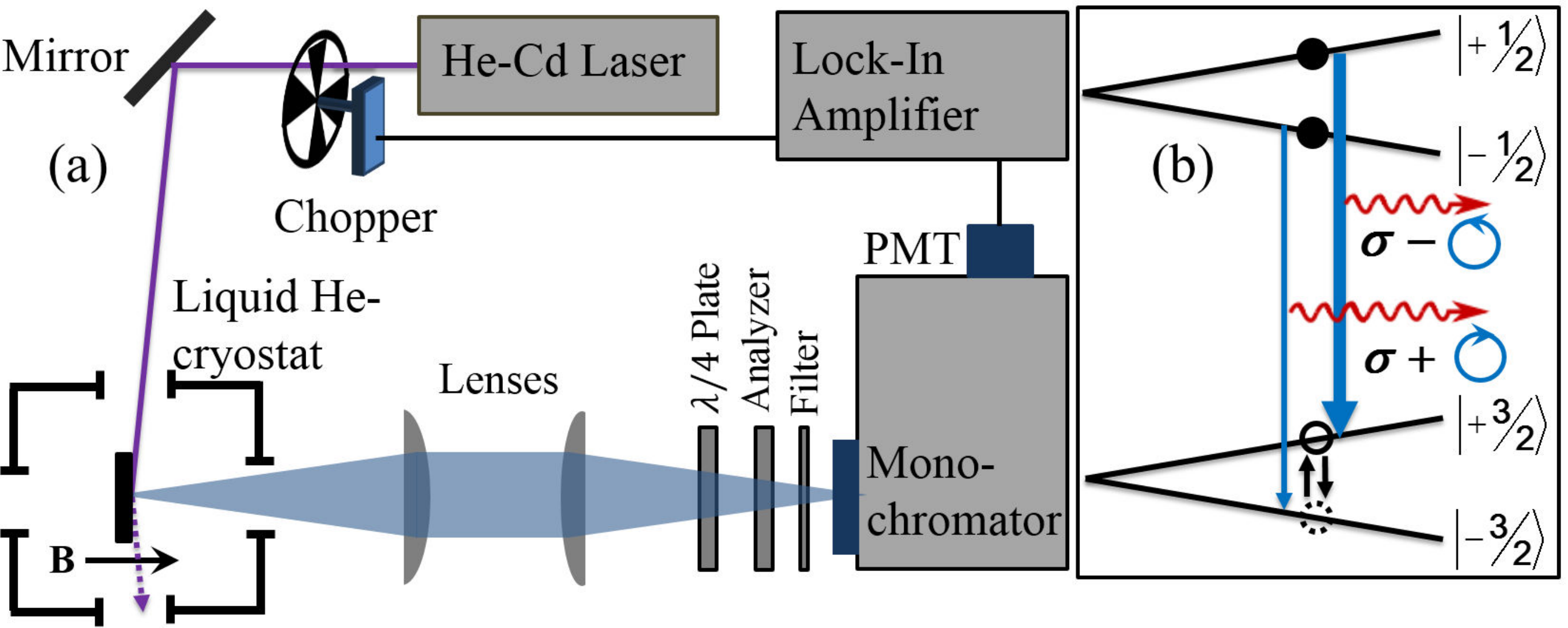}
		\caption{(a) Schematic depiction of the polarization selective magneto-photoluminescence (PL) spectroscopy setup. (b) Zeeman splitting of the conduction(valence) band minimum(maximum). Allowed optical transitions along with their polarization directions are also shown.}
		\label{fig1}
	\end{figure}

	GaN layers with different Gd concentrations were grown directly on 6H-SiC(0001) substrates using reactive molecular-beam epitaxy (RMBE) technique. A Gd-undoped GaN layer was also grown as the reference standard (sample R). One of the samples (sample E) was co-doped with both Gd and Si. More details about the growth can be found elsewhere\cite{dhar1a, dhar1b}.  Magnetization measurements show ferromagnetic like behavior even above 300~K in all of the Gd-doped samples. One of the Gd doped sample (sample D) was rapid thermally annealed at 600$^o$C for 30~s in flowing N$_2$ gas. See Tab.~\ref{tab1} to know more about these samples. A commercial grade hydride vapor phase epitaxy (HVPE) grown Si doped (N$_{Si}$ $\approx$ 1$\times$10$^{18}$~cm$^{-3}$) c-plane GaN(3~$\mu$m)/sapphire sample from TDI Inc. of USA, was used as another reference standard (Sample RS). Magneto-photoluminescence(PL) studies were carried out at 1.5~K in a liquid helium cryostat equipped with a split coil superconducting magnet and optical windows. Experimental setup is shown schematically in Fig.~\ref{fig1}(a). A linearly polarized He-Cd laser (325 nm) was used as the excitation source. Magnetic field was applied perpendicular to the sample surface (c-direction). Luminescence was collected along the magnetic field direction [Faraday geometry] through a 0.5~m focal length monochromator attached with a photomultiplier tube (PMT). Combination of an achromatic $\lambda/4$-plate and a Glan-Taylor calcite analyser was used to select either the $\sigma^+$ or $\sigma^-$ polarization of the luminescence.  Polarization of photoluminescence $P$ = $(I^+-I^-)/(I^++I^-)$, where $I^+$ and $I^-$ are the intensities associated with $\sigma^+$ and $\sigma^-$ polarized lights, respectively, was measured as a function of the applied magnetic field  at the peak positions of the defect related PL features. 

\begin{figure}
		\centering
		\includegraphics[width=1\linewidth]{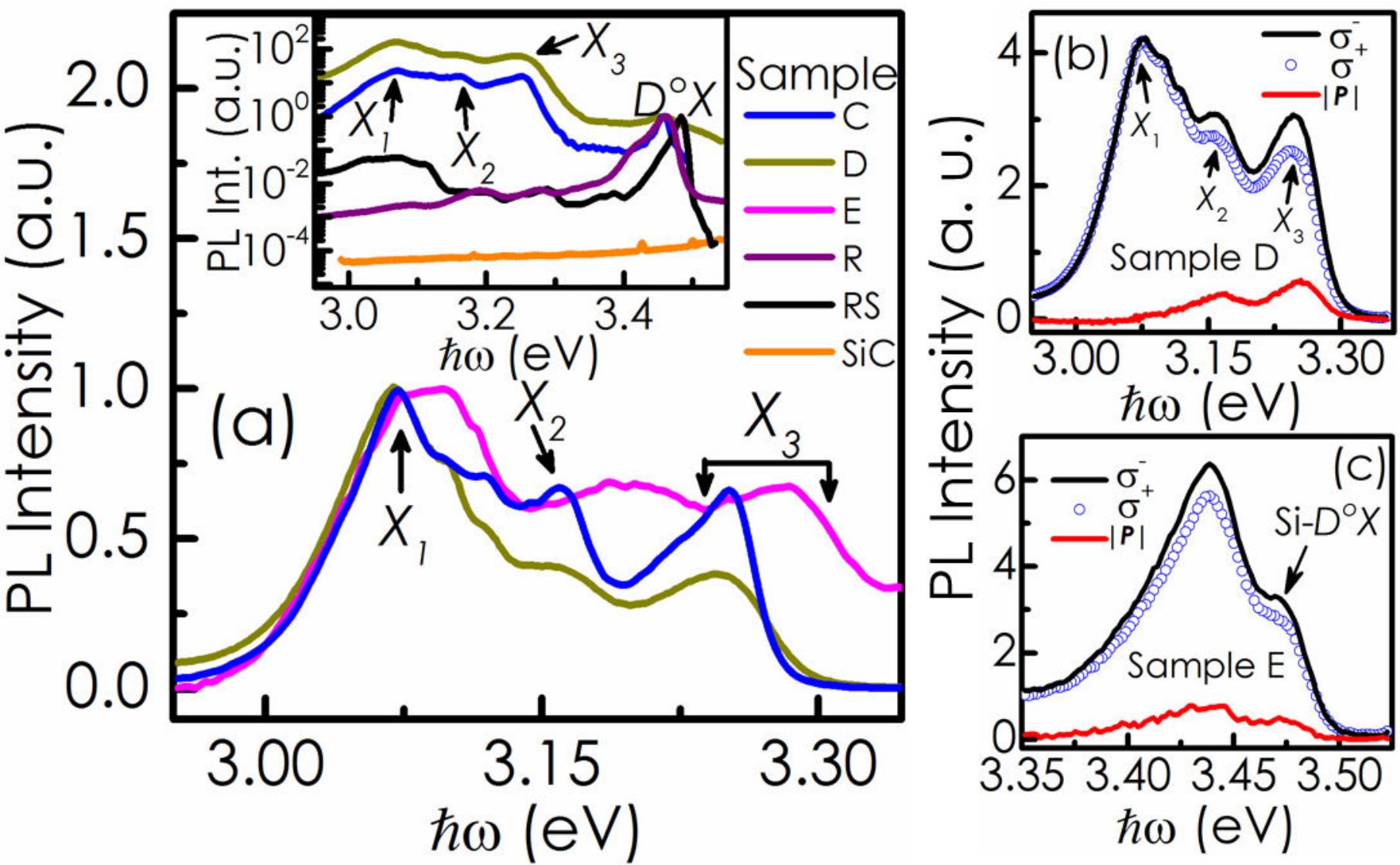}
		\caption{(a) PL spectra recorded at 1.5~K for samples with different Gd concentrations. Inset compares the PL spectra normalized with respect to the band edge transition for Gd-doped/undoped GaN samples and a piece of bare SiC substrate. PL spectra recorded at 1.5~K in $\sigma^+$ and $\sigma^-$ polarized states under a magnetic field of 7~T for (b) $X_1$, $X_2$, $X_3$ peaks in sample D and (c) $D^{o}X$ peaks in sample E. Red curves in panel (b) and (c) represent the magnitude of $P$ = $(I^+-I^-)/(I^++I^-)$ as a function of the photon energy.}
		\label{fig2}
	\end{figure}

 As shown in Figure~\ref{fig2}(a), 1.5~K PL spectra for both the reference samples R and RS is dominated by the neutral donor bound excitonic peak ($D^o X$) appearing at 3.457 and 3.48~eV, respectively. The high and low energy peaks can be attributed to $D^{o}X$ associated with Si and unintentionally incorporated oxygen shallow donors\cite{dhar1a,dhar1b}. PL spectra for Gd doped GaN samples C, D and E are featured by three relatively broad peaks appearing in the range of 3.05-3.1~eV ($X_1$), 3.15-3.2~eV ($X_2$) and 3.25-3.3~eV ($X_3$).  Bare SiC substrate does not show any feature in the photon energy range. Note that  $X_1$, $X_2$ and $X_3$ features, which are found to be present in all GaN:Gd samples, are attributed to certain defects produced in the GaN lattice as a result of Gd incorporation\cite{mishra2}. A close examination reveals that the broadening associated with these transitions increases while their peaks shift to higher energies as the concentration of Gd (N$_{Gd}$) increases. This implies an enhancement in the density of the three defect types with the Gd concentration. 
	
	As shown in Figure~\ref{fig2}(b), the degree of polarization is negative not only for the $D^o X$ transitions [see Figure~\ref{fig2}(c)] but also for all the $X$-features. This has indeed been found for all Gd doped samples investigated here. Note that the magnitude of PL polarization $|P|$ for $X_1$ peak is much weaker than those for $X_2$ and $X_3$ peaks[see Figure~\ref{fig2}(b)]. 
	
\begin{figure}
		\centering
		\includegraphics[width=0.9\linewidth]{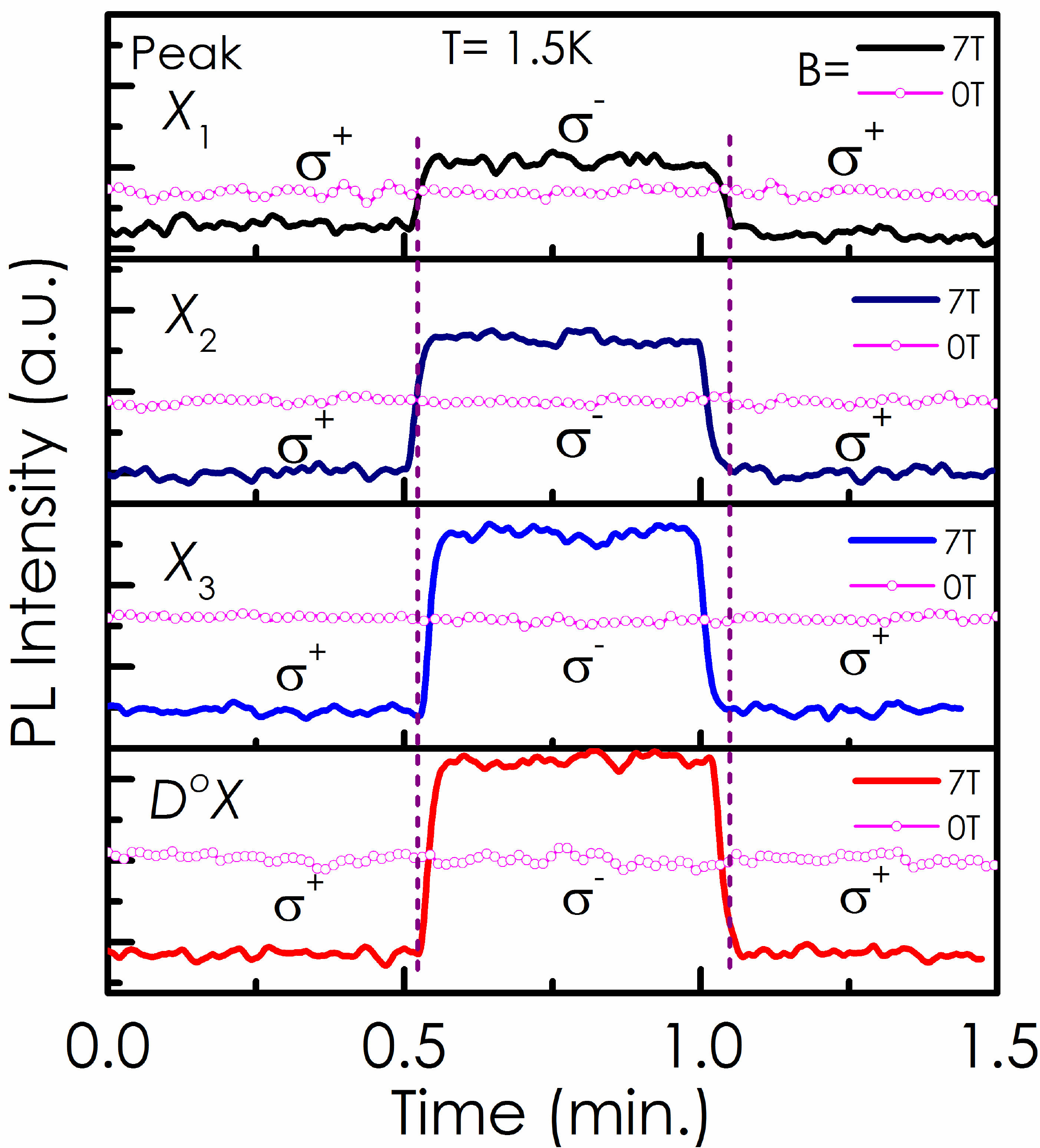}
		\caption{(a) PL intensities recorded at $X_1$, $X_2$, $X_3$ peaks for sample D and at $D^{o}X$ peak for sample RS at 0 and 7~T magnetic field as a function of time as the polarization selection is abruptly switched between $\sigma^+$ or $\sigma^-$ states at two time points.}
		\label{fig3}
	\end{figure}

	Fig.~\ref{fig3} compares the time dependent PL intensities recorded at $X$-features for sample E and at $D^{o}X$ peak for sample RS under zero and 7~T magnetic fields as the polarization selection is switched between $\sigma^+$ and $\sigma^-$ states at two time points. For all cases, the change in intensity due to the switching is evident only at non-zero fields.
	
	
 Since the $g$-factor for the conduction band $g_{e}(=1.95)$\cite{ge-factor} is comparable to that of 'A'-valence band holes $g_{h}(=2.25)$\cite{gh-factor} in GaN, the Zeeman splitting under a magnetic field along the c-axis is expected to be almost the same for the two bands. Note that a neutral donor bound exciton (NDBX) is composed of two electrons occupying $|+1/2>$ and $|-1/2>$ conduction band like J-states and a hole occupying either $|+3/2>$ or $|-3/2>$ 'A'-band like J-states. Photons collected along c-direction can either have $\sigma^+$ or  $\sigma^-$ polarizations as shown in Fig.~\ref{fig1}(b). In this geometry,  the intensity of PL for a NDBX transition with $\sigma^-$ polarization should be more as compared to that of $\sigma^+$ as $|+3/2>$ is energetically lower than $|-3/2>$ state for the holes. $P$ is thus negative for a NDBX transition. Observation of negative $P$ for all the $X$-features implies that these features are NDBX in nature as well.  
	
 At the steady state condition, as shown in the supplementary material\cite{suppl}, the polarization $P_i$ associated with $i$-th type of NDBX can be obtained as\cite{Kramer} $P_i = -(1-e^{-\Delta E_{h_i}(B)/k_B T})/(\gamma_i/\beta_{i}+1+e^{-\Delta E_{h_i}(B)/k_B T})$, where $\gamma_i$  and $\beta_i$ the rate of recombination and spin flip events, respectively, while $\Delta E_{h_i}$ the energy splitting of the two hole spin states associated with i-th type of NDBX. 
	
%
	

	It can be shown that  in this material, the term $e^{-\Delta E_{h_i}(B)/k_BT}$, in the expression of $P_i$, tends to zero at 1.5~K, for $B$ $>$ 1~T. Polarization $P_i$ can thus be written as $P_i$ = $-1/(\gamma_i/\beta_{i}+1)$ for $B$ $>$ 1~T, meaning the ratio $\gamma_i/\beta_{i}$ = 1/$|P|$ -1 can be obtained for Si-$D^o X$, and all the $X$-features, separately. Note that $\beta_{i}$ is expected to depend upon the magnetic field\cite{spin-dephasing}. However, $\gamma_i$ should be $B$-independent. If the i-th type of $X$-defects are ferromagnetically coupled, Zeeman splitting of the valence band like state of the exciton can be given by $\Delta E_{h_i}(B)$ = $g_{h_i}\mu_B B$ + $\alpha_i M(B)$, where $M(B)$ is the magnetization and $\alpha_i$ is the magnetic coupling coefficient. This means that each of the defects experiences an overall magnetic field of $B_T$ = $B$ + $B_{E}$. Saturation of magnetization $M_s$ leads to the saturation of the molecular field at $B_{Es}$ = $w_i M_s$, where $w_i$ the coupling constant for the defects. If $X_i$-feature is stemming from a region of ferromagnetically coupled i-th defects, where $B_{Es}$ prevails over $B$, $\beta_{i}$ is expected to show much weaker $B$-dependence beyond the saturation field.  Therefore, the ratio  $\gamma_i/\beta_{i}$ as a function of $B$ can carry the information about the involvement of individual defect types in ferromagnetic coupling.  
	
	\begin{figure}[hbtp]
		\centering
		\includegraphics[width=0.8\linewidth]{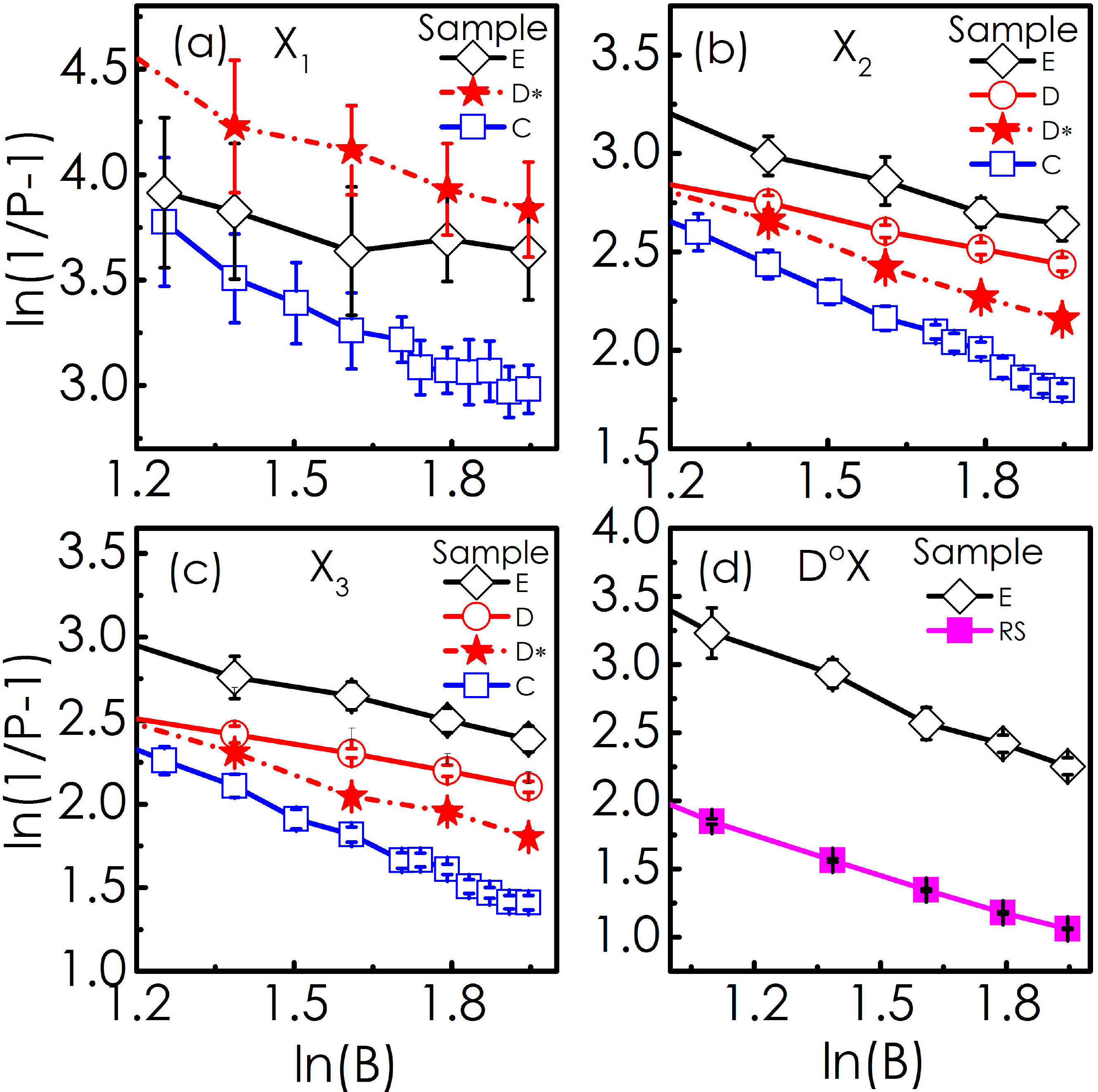}
		\caption{Plots of ln$(1/|P|-1)$ vs ln($B$) for (a) $X_1$, (b) $X_2$, (c) $X_3$ and (d) $D^{o}X$ peaks of the samples with different Gd concentrations ($N_{Gd}$). D* represents the sample D after annealing}
		\label{fig4}
	\end{figure}
	
	In Fig.~\ref{fig4}, ln$(1/|P|-1)$ is plotted as a function of $B$ for $X_1$ [Panel (a)], $X_2$ [Panel (b)], $X_3$ [Panel (c)] and $D^o X$ [Panel (d)] excitonic peaks in these samples.  Evidently, in all cases, the data show a linear variation at high fields, which suggests that $\beta_{i}$ $\propto$ $B^{\nu_i}$ for all these NDBX.  It is interesting to note that for all the $X$-features, the slope substantially decreases in samples with N$_{Gd}$ $>$ 10$^{17}$~cm$^{-3}$. Moreover, in sample D, the slope is increased for all the $X$-features upon annealing. Note that the saturation magnetization of the sample decreases by about 50$\%$ after annealing (see Tab.~\ref{tab1})\cite{mishra2}.  On the other hand, the slope for the $D^o X$ exciton in the Si doped reference sample RS is very much the same as that of the sample E, which is co-doped with both Si and Gd. These results thus suggest that the defects associated with $X$-features in Gd doped GaN samples must be experiencing an overwhelmingly large $B_{Es}$ field, meaning these defects are ferromagnetically coupled.  Interestingly, for the co-doped sample, $D^o X$ excitons associated with Si shallow donors do not experience any $B_{Es}$ field, which suggests that they are not involved in the ferromagnetic coupling. Role of the internal field in governing the slope of these plots becomes more explicit from the fact that upon annealing, $M_s$ (and hence $B_{Es}$) decreases and at the same time the slope increases in sample D.

\begin{figure}
		\centering
		\includegraphics[width=0.8\linewidth]{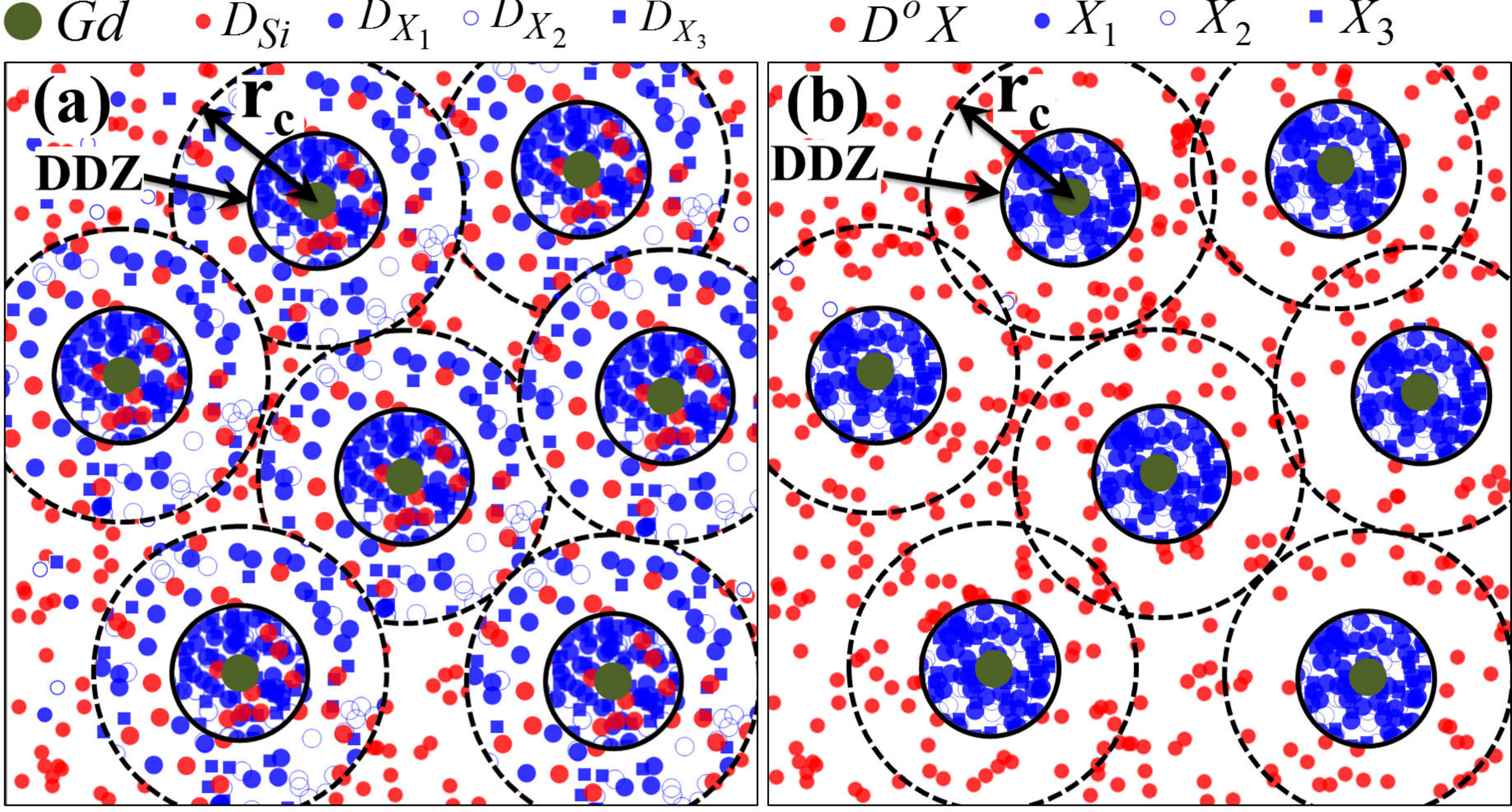}
		\caption{Schematic representation of (a) the distribution of Si shallow donors and the defects associated with $X_1$(defect $D_{X_1}$), $X_2$ (defect $D_{X_2}$) and $X_3$ (defect $D_{X_3}$) excitons, (b) the distribution of $D^{o}X$, $X_1$, $X_2$ and $X_3$ excitons.}
		\label{fig5}
	\end{figure}

	  Note that these defects are likely to be generated surrounding each Gd ion, whereas Si shallow donors are randomly distributed over the entire lattice as shown schematically in Fig.~\ref{fig5}(a). Close proximity of the defects in the regions surrounding the Gd sites leads to defect-defect ferromagnetic coupling \cite{mishra2}. This results in the formation of ferromagnetic domains surrounding every Gd site. Beyond a percolation threshold, a long range ferromagnetic order sets in. In this framework, the saturation magnetization can be expressed as M$_s$ = $p_{Gd}$N$_{Gd}$ + $p_o$N$_o$[1-$\exp$(-$v$N$_{Gd}$)], where $p_{Gd}$ and $p_o$ are the bare magnetic moments per Gd ion and defect, respectively, $N_o$ the defect density within the cluster and $v$=4/3$\pi r_c^3$ the volume of each cluster. In one of our earlier works, we have estimated  $p_o$N$_o$ = 4.68 $\times$ 10$^{19}$~$\mu_B$cm$^{-3}$  and $r_c$ = 22~nm by fitting the experimental data from Ref.1 with the above expression\cite{mishra2}. If an average magnetic moment $\approx$3~$\mu_B$(predicted for N-interstitials\cite{mitra}) is attributed per defect site,  N$_o$ comes out to be 1.56 $\times$ 10$^{19}$~cm$^{-3}$. It is noteworthy that a same order of magnitude of defect density has been estimated by Roever \etal in their MBE grown Gd:GaN samples\cite{Roever}. However, the actual distribution of N$_o$($r$) may not be uniform inside the defect sphere. Rather a reduction of N$_o$($r$) from the center to the surface is more realistic a scenario.  Note that in the co-doped sample (sample E), N$_{Si}$ is comparable with N$_o$. If radial variation of N$_o$ is taken into account, one can find a defect dominated zone (DDZ) around every Gd site, where  N$_o$ $>$ N$_{Si}$ as shown schematically in the figure.  $X$-excitons are thus expected to be mostly present in DDZs, whereas Si-$D^o X$ signal is arising from the regions outside these zones as depicted in Fig.~\ref{fig5}(b). Because of their non-involvement in ferromagnetic coupling, Si-donors do not experience any strong molecular field, even though they are located in the regions where the back ground $X$-defects are ferromagnetically coupled. This can explain why ln$(1/|P|-1)$ shows almost a $B$ independent behavior for all the $X$-features for this sample , while it decreases faster with increasing $B$ for the Si- $D^o X$ feature [see Fig.~\ref{fig4}]. It should be mentioned that even though the saturation magnetizations are comparable for sample D and C, variation of ln$(1/|P|-1)$ with $B$ is much faster in sample C than in sample D. The reason might be the strength of the molecular field  $w M_s$, where $w$ could depend upon the overlap of the defect clusters. Since in sample D,  N$_{Gd}$ is more than that in sample C, the overlap and hence $w M_s$ is expected to be higher. We believe that $w M_s$ prevails over $B$ in sample D, while in sample C, the field $B$ $>>$ $w M_s$.   
	  
	Observation of Fig.~\ref{fig4} suggests that the defects associated with all the X-features have neutral donor like states. Among all the defects, which are theoretically predicted to have magnetic moment, O$_i$,  V$_{Ga}$ and V$_N$-V$_{Ga}$ complexes do not contribute  any donor like state at a position matching with those of $X_1$, $X_2$ and $X_3$ \cite{Diallo, Wright}, meaning that the $D_{X_1}$, $D_{X_2}$ and $D_{X_3}$ defects are unlikely to be either of the three. In fact, Roever et al. using positron annihilation spectroscopy have shown that there is no direct correlation between Ga-vacancy and ferromagnetism observed in this material \cite{Roever}.  Nitrogen split interstitials, whose formation energy is one of the lowest among  all point defects and their complexes,  can have a 0/-1 state at about 0.48 eV below the conduction band minimum \cite{Diallo}. This position matches quite well with that of X$_1$ feature. Moreover, each N$_i$ is expected to contribute 3~$\mu_B$ of magnetic moment \cite{mitra}. Our finding, therefore, indicates that $D_{X_1}$ defects are N-interstitials. Positions of $X_2$ and $X_3$ do not match with those of any known point defect contributing neutral donor like states. It is noteworthy that upon annealing, PL intensity of $X_1$-feature reduces quite significantly as compared to those of $X_2$ and $X_3$ \cite{mishra2, mishra3} implying that $D_{X_2}$ and $D_{X_3}$ defects have better thermal stability than $D_{X_1}$ defects. It is plausible that $D_{X_2}$ and $D_{X_3}$ defects are also N$_i$s that make complexes with certain other point defects/impurities, which has better thermal stability than isolated N$_i$.
	   
	In conclusion, incorporation of Gd in GaN produce three types of donor like defects, which result in three neutral donor bound excitonic (NBDX) features appearing at about 3.05, 3.15 and 3.25~eV in the low temperature PL spectra. It has been shown that the dependence of spin-flip scattering rate on the magnetic field ($B$) for these NDBX features carry the information about the involvement of the associated defects in magnetic coupling. Our study shows that all the three signals must be stemming from those defects, which are ferromagnetically coupled with each other.  
	
	\section*{Acknowledgements}
	The authors acknowledge the financial support of this work by the Department of Science and Technology of the Government of India under the Project Code SR/S2/CMP-71/2012. We also would like to thank Paul-Drude-Institute, Berlin, Germany for the samples and Prof. S. Ghosh of TIFR, Mumbai for help in certain characterizations.


\bibliographystyle{apsrev4-1}
\bibliography{bib}

\end{document}